\documentstyle[twoside,psfig]{article}

\catcode`\@=11
\long\def\@makefntext#1{
\protect\noindent \hbox to 3.2pt {\hskip-.9pt  
$^{{\eightrm\@thefnmark}}$\hfil}#1\hfill}		

\def\thefootnote{\fnsymbol{footnote}}
\def\@makefnmark{\hbox to 0pt{$^{\@thefnmark}$\hss}}	
	
\def\ps@myheadings{\let\@mkboth\@gobbletwo
\def\@oddhead{\hbox{}
\rightmark\hfil\eightrm\thepage}   
\def\@oddfoot{}\def\@evenhead{\eightrm\thepage\hfil
\leftmark\hbox{}}\def\@evenfoot{}
\def\sectionmark##1{}\def\subsectionmark##1{}}

\oddsidemargin=\evensidemargin
\renewcommand{\thefootnote}{\fnsymbol{footnote}}

\newcounter{sectionc}\newcounter{subsectionc}\newcounter{subsubsectionc}
\renewcommand{\section}[1] {\vspace{12pt}\addtocounter{sectionc}{1} 
\setcounter{subsectionc}{0}\setcounter{subsubsectionc}{0}\noindent 
	{\tenbf\thesectionc. #1}\par\vspace{5pt}}
\renewcommand{\subsection}[1] {\vspace{12pt}\addtocounter{subsectionc}{1} 
	\setcounter{subsubsectionc}{0}\noindent 
	{\bf\thesectionc.\thesubsectionc. {\kern1pt \bfit #1}}\par\vspace{5pt}}
\renewcommand{\subsubsection}[1] {\vspace{12pt}\addtocounter{subsubsectionc}{1}
	\noindent{\tenrm\thesectionc.\thesubsectionc.\thesubsubsectionc.
	{\kern1pt \tenit #1}}\par\vspace{5pt}}
\newcommand{\nonumsection}[1] {\vspace{12pt}\noindent{\tenbf #1}
	\par\vspace{5pt}}

\newcounter{appendixc}
\newcounter{subappendixc}[appendixc]
\newcounter{subsubappendixc}[subappendixc]
\renewcommand{\thesubappendixc}{\Alph{appendixc}.\arabic{subappendixc}}
\renewcommand{\thesubsubappendixc}
	{\Alph{appendixc}.\arabic{subappendixc}.\arabic{subsubappendixc}}

\renewcommand{\appendix}[1] {\vspace{12pt}
        \refstepcounter{appendixc}
        \setcounter{figure}{0}
        \setcounter{table}{0}
        \setcounter{lemma}{0}
        \setcounter{theorem}{0}
        \setcounter{corollary}{0}
        \setcounter{definition}{0}
        \setcounter{equation}{0}
        \renewcommand{\thefigure}{\Alph{appendixc}.\arabic{figure}}
        \renewcommand{\thetable}{\Alph{appendixc}.\arabic{table}}
        \renewcommand{\theappendixc}{\Alph{appendixc}}
        \renewcommand{\thelemma}{\Alph{appendixc}.\arabic{lemma}}
        \renewcommand{\thetheorem}{\Alph{appendixc}.\arabic{theorem}}
        \renewcommand{\thedefinition}{\Alph{appendixc}.\arabic{definition}}
        \renewcommand{\thecorollary}{\Alph{appendixc}.\arabic{corollary}}
        \renewcommand{\theequation}{\Alph{appendixc}.\arabic{equation}}
        \noindent{\tenbf Appendix \theappendixc #1}\par\vspace{5pt}}
\newcommand{\subappendix}[1] {\vspace{12pt}
        \refstepcounter{subappendixc}
        \noindent{\bf Appendix \thesubappendixc. {\kern1pt \bfit #1}}
	\par\vspace{5pt}}
\newcommand{\subsubappendix}[1] {\vspace{12pt}
        \refstepcounter{subsubappendixc}
        \noindent{\rm Appendix \thesubsubappendixc. {\kern1pt \tenit #1}}
	\par\vspace{5pt}}

\newcommand{\textlineskip}{\baselineskip=13pt}
\newcommand{\smalllineskip}{\baselineskip=10pt}

\def\eightcirc{
\begin{picture}(0,0)
\put(4.4,1.8){\circle{6.5}}
\end{picture}}
\def\eightcopyright{\eightcirc\kern2.7pt\hbox{\eightrm c}} 

\newcommand{\copyrightheading}[1]
	{\vspace*{-2.5cm}\smalllineskip{\flushleft
	{\footnotesize Modern Physics Letters A, #1}\\
	{\footnotesize $\eightcopyright$\, World Scientific Publishing
	 Company}\\
	 }}

\def\abstracts#1#2#3{{
	\centering{\begin{minipage}{4.5in}\footnotesize\baselineskip=10pt
	\parindent=0pt #1\par 
	\parindent=15pt #2\par
	\parindent=15pt #3
	\end{minipage}}\par}} 

\renewenvironment{thebibliography}[1]
	{\frenchspacing
	 \ninerm\baselineskip=11pt
	 \begin{list}{\arabic{enumi}.}
        {\usecounter{enumi}\setlength{\parsep}{0pt}     
	 \setlength{\leftmargin 12.7pt}{\rightmargin 0pt} 
         \setlength{\itemsep}{0pt} \settowidth
	{\labelwidth}{#1.}\sloppy}}{\end{list}}

\newcounter{itemlistc}
\newcounter{romanlistc}
\newcounter{alphlistc}
\newcounter{arabiclistc}

\newcommand{\fcaption}[1]{
        \refstepcounter{figure}
        \setbox\@tempboxa = \hbox{\footnotesize Fig.~\thefigure. #1}
        \ifdim \wd\@tempboxa > 5in
           {\begin{center}
        \parbox{5in}{\footnotesize\smalllineskip Fig.~\thefigure. #1}
            \end{center}}
        \else
             {\begin{center}
             {\footnotesize Fig.~\thefigure. #1}
              \end{center}}
        \fi}

\newcommand{\tcaption}[1]{
        \refstepcounter{table}
        \setbox\@tempboxa = \hbox{\footnotesize Table~\thetable. #1}
        \ifdim \wd\@tempboxa > 5in
           {\begin{center}
        \parbox{5in}{\footnotesize\smalllineskip Table~\thetable. #1}
            \end{center}}
        \else
             {\begin{center}
             {\footnotesize Table~\thetable. #1}
              \end{center}}
        \fi}

\def\@citex[#1]#2{\if@filesw\immediate\write\@auxout
	{\string\citation{#2}}\fi
\def\@citea{}\@cite{\@for\@citeb:=#2\do
	{\@citea\def\@citea{,}\@ifundefined
	{b@\@citeb}{{\bf ?}\@warning
	{Citation `\@citeb' on page \thepage \space undefined}}
	{\csname b@\@citeb\endcsname}}}{#1}}

\newif\if@cghi
\def\cite{\@cghitrue\@ifnextchar [{\@tempswatrue
	\@citex}{\@tempswafalse\@citex[]}}
\def\citelow{\@cghifalse\@ifnextchar [{\@tempswatrue
	\@citex}{\@tempswafalse\@citex[]}}
\def\@cite#1#2{{$\null^{#1}$\if@tempswa\typeout
	{IJCGA warning: optional citation argument 
	ignored: `#2'} \fi}}

\def\pmb#1{\setbox0=\hbox{#1}
	\kern-.025em\copy0\kern-\wd0
	\kern.05em\copy0\kern-\wd0
	\kern-.025em\raise.0433em\box0}

\def\fnm#1{$^{\mbox{\scriptsize #1}}$}
\def\fnt#1#2{\footnotetext{\kern-.3em
	{$^{\mbox{\scriptsize #1}}$}{#2}}}

\def\fpage#1{\begingroup
\voffset=.3in
\thispagestyle{empty}\begin{table}[b]\centerline{\footnotesize #1}
	\end{table}\endgroup}

\font\tenrm=cmr10
\font\tenit=cmti10 
\font\tenbf=cmbx10
\font\bfit=cmbxti10 at 10pt
\font\ninerm=cmr9

\font\eightrm=cmr8

\def\qed{\hbox{${\vcenter{\vbox{			
   \hrule height 0.4pt\hbox{\vrule width 0.4pt height 6pt
   \kern5pt\vrule width 0.4pt}\hrule height 0.4pt}}}$}}

\renewcommand{\thefootnote}{\fnsymbol{footnote}}	

\def \ba {\begin{array}}
\def \ea {\end{array}}

\def \bea {\begin{eqnarray}}
\def \eea {\end{eqnarray}}

\def \be {\begin{equation}}
\def \ee {\end{equation}}

\def\nn{\nonumber}

\begin{document}
\setlength{\textheight}{7.7truein}  


\normalsize\textlineskip
\thispagestyle{empty}
\setcounter{page}{1}

\copyrightheading{}			

\vspace*{0.88truein}

\fpage{1}
\centerline{\bf AN EFFECTIVE QUANTUM MECHANISM FOR MASS GENERATION}
\baselineskip=13pt
\centerline{\bf IN DIFFEOMORPHISM-INVARIANT THEORIES\footnote{Work partially
supported by the DGICYT.}}
\vspace*{0.37truein}
\centerline{\footnotesize J. L. JARAMILLO}
\baselineskip=12pt
\centerline{\footnotesize V. ALDAYA}
\vspace*{10pt}
\centerline{\footnotesize\it Instituto Astrof\'\i sica Andaluc\'\i a
(CSIC), Apartado Postal 3004}
\baselineskip=10pt
\centerline{\footnotesize\it Granada 18080, Spain}
\baselineskip=12pt
\centerline{\footnotesize\it Instituto Carlos I de F\'\i sica Te\'orica y 
Computacional, Facultad de Ciencias,} 
\baselineskip=10pt
\centerline{\footnotesize\it Universidad de Granada, Campus de 
Fuentenueva}
\baselineskip=10pt
\centerline{\footnotesize\it Granada 18002, Spain}
\vspace*{0.225truein}

\vspace*{0.225truein}
\vspace*{0.21truein}
\abstracts{We propose a scenario for particle-mass generation, assuming the 
existence of a physical regime where, firstly, physical particles can be 
considered as point-like objects moving in a background space-time and, 
secondly, their mere presence spoils the invariance under 
the local diffeomorphism group, resulting in an anomalous realization of the 
latter. Under these hypotheses, 
we describe mass generation starting from the massless free theory. The 
mechanism is not sensitive to the detailed description of the underlying
theory at higher energies, leaning only on 
general structural features of it, specifically diffeomorphism invariance.  
}{}{}


\textlineskip			
\vspace*{12pt}			

\vspace*{-0.5pt}

\setcounter{footnote}{0}
\renewcommand{\thefootnote}{\alph{footnote}}

\noindent
The problem we address in this work is the origin of particle masses. Even 
though a strong emphasis has been placed on this issue
throughout the development of modern physics, the subject seems far from being 
resolved. 

The correction of the mass of a particle as an effective consequence of its 
interaction with the surrounding environment is a very old idea that can be 
tracked to nineteenth-century hydrodynamics. In fact, for many different
physical systems describing the motion of an object inside a 
classical continuum
fluid, the solution of hydrodynamical equations admits an effective
treatment in which the object behaves as in free motion with a corrected or 
{\it renormalized} mass which depends on general features such as boundary 
conditions.  The extension of these
ideas to electrodynamics led J.J.Thomson to the introduction of the notion
of {\it electromagnetic mass} of a charge as a consequence of the interaction 
with its own electromagnetic field, fundamental element in the later Lorentz's 
theory of the 
electron \cite{Lorentz}. With the arrival of the Quantum Theory, the efforts 
by Kramers 
(largely inspired in Lorentz's insights) resulted in the connection between 
the previous classical 
ideas and the new problems related to the divergences appearing in the 
calculation of the electron self-energy, leading to a radiative 
mass renormalization. In the early days of Quantum Electrodynamics
there was hope of developing a fundamental theory that would
eliminate the divergences and successfully derive the actual values of its
characteristic parameters. However, the eventual resolution of the problem
by implementing the renormalization program finally led
to a situation in which Quantum Field Theory (QFT) appears as
an effective theory. In fact, physics beyond a certain energy scale is 
not probed, though the renormalization of certain parameters of
the model, among them the masses of the particles, non-trivially affects  
lower-energy physics. In this scenario, the idea 
of mass as self-energy has withered away to moot status.
Nevertheless, conceptually different mechanisms can be devised for addressing 
partial yet fundamental aspects of the mass-origin problem, an example of 
which is the use of lattice QCD techniques for light hadrons \cite{Wilczek}. 
In any case, questions such as lepton 
masses or cosmological {\it dark matter}, remain open.
 
From this historical detour (see \cite{renormalization} for further details), 
we extract our two main guidelines.
Firstly, we take up the old idea of emphasizing the
interaction with the surrounding fields as fundamental in the generation 
of mass and, secondly, we adopt an effective aproach in which physics beyond a 
certain scale is not discussed.  
The presence of unanswered questions suggests
the introduction of physics often ignored in the mass
generation problem. An appealing (and obvious) candidate for the 
missing physical ingredient is Gravity with its associated diffeomorphism 
invariance, generally not considered in high-energy particle physics. 
Therefore, the only explicit condition we shall impose on the underlying 
fundamental theory is an essential role for the notion of diffeomorphism 
invariance.

When adopting the above-mentioned effective attitude, it seems reasonable 
to admit the existence of a 
scale of energies in which standard QFT is a good approximation, and 
its notion 
of a particle as a local excitation of the vacuum resulting from the action 
of a local field operator applies. We are also implicitly accepting a notion 
of space-time as a differentiable manifold making up the background in 
which particles move. We shall phenomenologically separate the intrinsic 
dynamics of this effective background, governed by classical General 
Relativity, from the effect that the underlying diffeomorphism 
invariance could exert in the quantum proccess of particle creation.

We are therefore studying a regime in which physical particles
can be considered as point-like objects and the classical 
dynamics of space-time is decoupled (adiabatic condition).
The adjective {\it physical} appearing here is essential, as opossed to 
the {\it ideal} test particles, causing an 
effective breakdown of the space-time notion at the point itself on which 
the particle lies. 
We are suggesting that physical particles literally {\it pierce} 
space-time, producing a hole. This has profound consequences in the 
quantum model describing particle creation. It can be shown \cite{Pressley}
that the presence of a hole in a two-dimensional manifold induces anomalous
(central) terms in the quantum commutators between (some of) the generators 
of diffeomorphism invariance, thus spoiling this classical symmetry 
(even though this can be properly healed in specific theories). We propose
that this phenomenon generalizes
to realistic space-time dimensions, inducing an anomalous realization
of classical diffeomorphism symmetry in the effective quantum process of 
particle creation, something that could be supported by the analysis of the 
leading terms of appropiate Operator Product Expansions. This does not 
contradict an exact implementation of 
this symmetry at higher energies, when using a more fundamental model for
the coupling between the gravitational and matter degrees of freedom. It 
simply means that the price we must pay for avoiding such a detailed 
description, and admitting an effective treatment in which classical 
space-time is decoupled, is the acceptance of
a breakdown of classical diffeomorphism gauge invariance.

The previous heuristic motivations can be  
synthesized in the following hypothesis: there exists an effective regime in 
which physical particles are point-like and their creation process entails
a breakdown of classical diffeomorphism invariance, the latter being realized 
in an anomalous way.

The presence of an anomaly in a local gauge theory obstructs the reduction 
of degrees of freedom for which the gauge symmetry is devised, entailing an
enlargement of the physical phase space performed by the spurious 
(in principle) modes \fnm{a}\fnt{a} {A familiar example of this 
phenomenon in string theory 
is the Liouville mode in the non-critical string.}. This issue poses serious 
concerns for the 
consistency of the theory, at least when applying standard techniques, 
something especially critical when 
addressing the gauge theory as  fundamental (attempts 
to construct consistent anomalous theories do exist \cite{Fadeev} and the 
above-mentioned 
explicit appearance of extra degrees of freedom can be made apparent). 
However, the presence of an anomaly can also be 
interpreted as a signature for understanding the theory as a low-energy
effective model, indicating the existence of new physics at higher energies 
\cite{Preskill}. This is precisely the situation we are dealing with here.
The influence of higher-energy degrees of freedom 
is encoded in some effective degrees of freedom arising in the anomalous 
low-energy theory.

A standard way in which an anomaly manifests itself, in accordance
with the considerations above, is through the 
appearance of extra terms in the quantum commutators with respect to the 
ones defining the classical symmetry. Therefore, we propose that the 
diffeomorphism symmetry is realized in the
effective theory as an extension (not necessarily central) of the classical
local diffeomorphism algebra. For concreteness we focus on the tensorial
extensions, which are classified in \cite{Dhu} and discussed in 
\cite{Larsson}. We shall work in momentum space and denote the diffeomorphism 
generators by $\hat{L}_\mu({\bf m})$, the fields corresponding 
to the particles generically by $\hat{\Phi}_a({\bf m})$ and the tensorial
extensions by $\hat{A}_i({\bf m})$ ($\mu$ is a space-time index, $a$ and $i$ 
internal indices and  
${\bf m}$ a vector labelling momentum space). In this notation, the quantum 
brackets are given by:
\bea
\left[\hat{L}_\mu({\bf m}), \hat{L}_\nu({\bf n})\right]&=&
m_\nu \hat{L}_\mu({\bf m+n})-
n_\mu \hat{L}_\nu({\bf m+n})+c^i_{\mu\nu}({\bf m},{\bf n})
\hat{A}_i({\bf m+n}) \nn \\
\left[\hat{L}_\mu({\bf m}),\hat{\Phi}_a({\bf n})\right]&=&-n_\mu
\hat{\Phi}_a({\bf m+n}) \nn \\
\left[\hat{\Phi}_a({\bf m}),\hat{\Phi}_b({\bf n})\right]&=&
\hat{\alpha}_{ab}({\bf m},{\bf n})
\ \ ,  \label{algc}
\eea
where $c^i_{\mu\nu}({\bf m},{\bf n})$ is the cocycle linked to the anomalous 
extension giving
effective dynamical content to the diffeomorphisms and 
$\hat{\alpha}_{ab}({\bf m},{\bf n})$ provides the standard commutators 
of the free fields.

To give a specific meaning to the entire foregoing discussions, we
need to  construct explicitly a physical model describing  
dynamics consistent with the algebra (\ref{algc}). A particularly well 
suited formalism for such a task is the so-called
Group Approach to Quantization (GAQ, see \cite{GAQ}). In short, the 
main achievement of this approach is the construction of physical dynamics
out of a given Lie algebraic structure taken as the only physical input. 
The technique, in some points, resembles Kirillov's construction of 
dynamics on the coadjoint orbits of a group \cite{Kirillov} and shares 
some important general features with Geometric Quantization \cite{GC}.
The final outcome is an explicit unitary and irreducible representation
of the operators in the starting Lie algebra.

When applying these techniques to algebras of the type (\ref{algc}),
we obtain maximum-weight representations (possessing a unique vacuum in the 
Hilbert space), where the corresponding diffeomorphism operators 
$\hat{L}_\mu({\bf m})$ act and genuinely raise and lower the physical states,
according to their gained dynamical content. A most important precise 
and general (perturbative \fnm{b}\fnt{b}{A crucial step of GAQ consists
in exponentiating the starting Lie algebra. When the latter is 
infinite-dimensional
this is a enormous task and only an order-by-order procedure is generally 
feasible, leading to perturbative though renormalized results.}) result is 
the construction of an 
(effective) Hamiltonian operator for the system with the general form:
\bea
\hat{H}_{eff}&=&\hat{H}_{free}\left(\hat{\Phi}^\dagger,\hat{\Phi}\right) + 
\sum_{\bf m}\theta^{\mu\nu}({\bf m})(\hat{L}_{\mu})^\dagger({\bf m})
\hat{L}_{\nu}({\bf m})+ \nn \\
&+&\hat{H}_{mixing}\left(\hat{\Phi}^\dagger,\hat{\Phi},
(\hat{L}_{\mu})^\dagger,\hat{L}_{\mu}\right) \ \ , \label{hamiltoniano}
\eea
where $\hat{H}_{free}$ is the Hamiltonian corresponding to the free massless
field theory, the second term is a pure dynamical-diffeomorphism quadratic 
contribution to the energy ($\theta^{\mu\nu}({\bf m})$ is a {\it c-number}
function on ${\bf m}$ which closely depends on the inverse of the cocycle 
$c^i_{\mu\nu}({\bf m},{\bf n})$) and $\hat{H}_{mixing}$ corresponds to 
higher-power terms involving a potential mixing among all the operators.
Appearing perhaps as an odd phenomenon, the lowest-order 
term producing interaction is not found inside $\hat{H}_{mixing}$, but 
already in the quadratic diffeomorphism one, the reason being the
non-canonical form of the commutators in (\ref{algc}), in particular the
second one. This will be apparent in a specific example below.
Regarding the expression (\ref{hamiltoniano}), our claim is that the terms 
correcting the free Hamiltonian, could account for the mass terms in the
effective theory. 

Finally, we are in the position of unambiguosly stating our conjecture:
a crucial part of mass generation can be phenomenologically described as 
a (radiative) correction resulting from the interaction between the massless
fields and some effective degrees of freedom appearing from the mere existence 
of particles.

Of course, a real prediction of this contribution to the mass would require 
a knowledge of the underlying
fundamental theory, since it plays the role of fixing the values of the 
extensions in the algebra (\ref{algc}) and therefore of the key 
$\theta^{\mu\nu}({\bf m})$. Beyond that, the mechanism is not sensitive 
to the  higher-energy detailed description which could find support on 
strings theory, loop quantum gravity, a non-commutative version of geometry,
a more sophisticated QFT or another effective yet more fundamental model,
such as worm-holes (see \cite{Carlos}) in Euclidean quantum gravity. A 
serious attempt to provide a realistic example in 
this context deserves a careful analysis on the potential 
anomalous breaking of diffeomorphism invariance in current candidates for
fundamental theories. For the time being, we present an over-simplified 
illustrative example, consisting
of a free real scalar field in one (compact) spatial dimension. With the 
ansatz that only the spatial diffeomorphisms become dynamical, the
relevant algebra is:
\bea
\left[\hat{L}_m, \hat{L}_n\right]&=&(m-n)\hat{L}_{m+n}+ c m^3 \delta_{m+n,0}
\nn\\
\left[\hat{L}_m,\hat{a}_n\right]&=&-n \hat{a}_{m+n} \nn \\
\left[\hat{a}_m,\hat{a}_n\right]&=&m\delta_{m+n,0} \ \ .  \label{algescal} 
\eea
The Hilbert space is constructed from a unique vacuum state $|0\!>$, by 
applying the
creation operators $\hat{\alpha}^\dagger_n\equiv\frac{1}{\sqrt{n}}
\hat{a}_{-n}$ and $\hat{L}^\dagger_{n}\equiv\hat{L}_{-n}$, with $n\!>\!0$, and 
where the annihilation
operators are given by $\hat{\alpha}_n\equiv\frac{1}{\sqrt{n}}
\hat{a}_{n}$ and $\hat{L}_{n}$ ($n\!\geq\!0$). The operators 
$\hat{\alpha}^\dagger_n$, $\hat{\alpha}_n$ now satisfy the standard
canonical commutators: $[\hat{\alpha}_n,\hat{\alpha}^\dagger_m]=\delta_{n,m}$.
The perturbative calculation of the Hamiltonian of the system (which can be 
derived from Noether invariants in \cite{primeros}, together with a proper
setting in the values of the central extensions there) yields:
\bea
\hat{H}=\sum_{n>0}\left(n\hat{\alpha}^\dagger_n\hat{\alpha}_n+\frac{1}{c n^2}
\hat{L}^\dagger_{n}\hat{L}_{n}+...\right) \ \ .\label{ham} 
\eea
The expression from the standard treatment of a free field with mass $M$  
presents only a term $\hat{H}_{field}=\sum_{n>0}\;\sqrt{M^2+n^2}
\hat{\alpha}^\dagger_n
\hat{\alpha}_n$ (we are explicitly omitting zero-energy terms). Taking into 
account that the expression in (\ref{ham}) is 
only perturbative, we must look for a regime in which we can coherently 
compare it
with $\hat{H}_{field}$\fnm{c}\fnt{c}{The comparison with the corresponding 
expression in a first-quantization
theory such as the standard treatment of the bosonic string would be formally
more straightforward, but this is not our aim here.}. This can be achieved 
by expanding the field 
dispersion relation for large $n$. Thus we have 
\bea
\hat{H}_{field}=\sum_{n>0}n
\sqrt{1+\frac{M^2}{n^2}}\hat{\alpha}^\dagger_n
\hat{\alpha}_n=\sum_{n>0}\left(n\hat{\alpha}^\dagger_n\hat{\alpha}_n+
\frac{M^2}{2n}\hat{\alpha}^\dagger_n\hat{\alpha}_n+...\right) \ \ .
\label{hamfield}
\eea
The explicit way of comparing both approaches is accomplished by evaluating 
(\ref{ham}) and (\ref{hamfield}) 
on physical states, in particular on 1-particle states
$|k\!>= \hat{\alpha}^\dagger_k|0\!>$. 
One can raise the question about the convenience of using 
$|k\!>= \hat{\alpha}^\dagger_k|0\!>$ to 
implement the physical 1-particle states in the theory with dynamical 
diffeomorphisms, where the excitation of these effective modes would suggest 
the possibility
of a more general linear combination containing $\hat{L}^\dagger_k|0\!>$
states. At worst, one could consider (\ref{ham}) in the spirit of  
perturbation theory over a free massless scalar field, thus using  
the non-perturbated $|k\!>= \hat{\alpha}^\dagger_k|0\!>$ to evaluate
first-order corrections to the energy levels. This is the approach we shall 
use here. 

The correction to the excitation energy of the massless particle,
calculated in the quantum effective theory
derived from (\ref{algescal}), accounts for the energy of the interaction 
with the 
effective diffeomorphism degrees of freedom. Even though this energy could 
show a complicated behaviour in the momentum of the particle, we separate
the intermediate ({\it low}) and very high-momenta dependence and attempt  
to extract the energy related to the mass out of the form of the 
interaction energy at the lowest appearing  momenta.   
Therefore, when evaluting the second term in (\ref{ham}) (which we shall 
denote by $\hat{H}_L$) we expect to find 
an expression that can be identified with the one coming from the second term 
in (\ref{hamfield}) plus an energy corresponding to very high-momenta 
dependence ($E_{h-m}$):
\bea
\frac{<\!k|\hat{H}_L|k\!>}{<\!k|k\!>}=\frac{M^2}{2k}+E_{h-m} \ \ .
\eea
The evaluation of the first member gives the finite result:
\bea
\frac{<\!k|\hat{H}_L|k\!>}{<\!k|k\!>}=\frac{1}{c}\sum_{n=1}^{k-1}
\frac{k(k-n)}{n^2} \ \ . \label{correccion}
\eea
The identification of the particle-mass out of (\ref{correccion}) is 
guaranteed by the existence of terms in the sum with a $k^{-1}$ behaviour
(using large $k$). Therefore,
\bea
\frac{M^2}{2k}\sim\frac{1}{ck}\Rightarrow M\sim\sqrt{\frac{2}{c}} \ \ .
\eea

The rest of the energy is in fact a very high-momenta correction to the 
interacting
energy, implying high-energy deviations from the dispersion relation we are 
trying to match. The inclusion of these corrections in a corresponding field 
theory without
diffeomorphisms, would imply the correction of the propagator of the free 
field. Therefore, as long as we do not deal with very high energies, a field
theory of a Klein-Gordon scalar field with 
mass $M=\sqrt{\frac{2}{c}}$ is a good model of the effective quantum theory 
defined by (\ref{algescal}), thus involving a rationale for the origin of the 
mass 
parameter in the field theory. However, when we extend the application range
of the Klein-Gordon model beyond its limits, we are disregarding the 
above mentioned high-momentum interacting energy and, as a matter of fact, 
we are decoupling it from
the non-gravitational interactions. We have then an energy, and therefore a 
source for the gravitational field dynamics described by Einstein equations
(which we decoupled from the very beginning) that is not {\it seen} by the rest
of physical interactions,
and thus it is completely dark. It is tempting to suggest speculatively 
this as an avenue towards the dark-matter problem. 
When studying the two-particle states, we find
\bea
\frac{<\!kl|\hat{H}_L|kl\!>}{<\!kl|k\!l>}&=&\frac{k}{c}\sum_{n=1}^{k-1}
\frac{k-n}{n^2} + \frac{l}{c}\sum_{n=1}^{l-1}
\frac{l-n}{n^2}+ \nn \\
&+& \frac{kl}{c(1+\delta_{n,k})}\left(\frac{1}{(k+l)^2}+
\frac{\delta_{k\neq l}}{(k-l)^2}\right) \ \ ,
\eea
where the first two terms in r.h.s. correspond to the masses of the particles 
and its high-momenta corrections while the third term can be interpretated as 
an extra energy needed to maintain the particles {\it separated} (note that 
it is positive, so that we must do some work to have separated particles).

\noindent Finally, we should point out that the use of the 
Virasoro algebra does not turn the two-dimensional example into a too special 
case, since a direct generalization to higher dimensions is in fact provided 
by the non-central Virasoro-like extension\cite{Larsson}
obtained by  making
$c^i_{\mu\nu}({\bf m,n})=c\; m_\mu n_\nu(m^i-n^i)$ and $\hat{A}_i({\bf m})=
\hat{S}_i({\bf m})$
in (\ref{algc}), although involving much more cumbersome expressions.
The only aim of the presented example is that of providing a taste of and 
sparking intuition for
the potentialities of expressions (\ref{hamiltoniano}) and (\ref{ham}), the 
real point we wish to emphasize. Rigorous analyses require a subtle weaving 
together of the possible extensions in (\ref{algc}), the field dispersion 
relation we seek to fit and the possibility of more complex settings in 
which those elements may act. 

In conclusion, we have posed a simple framework for mass 
generation
of particles by proposing a mechanism capable of endowing massless free fields 
with a non-zero mass. Even if this is a tiny effect, it would provide a 
non-zero germ suitable of being amplified with other mechanisms such as the 
multiplicative renormalization appearing in QED for the electron. The present 
description has an effective nature and is quite
insensitive to the underlying fundamental theory, provided that 
diffeomorphism invariance plays a fundamental role. This last point is 
reinforced by Mach's conceptual intuitions linking inertia and Gravity. 
A heavy use of the image of particle creation as an inherently
quantum process is displayed. This can be explicitly seen in the example 
discussed by underlining the dependence of particle masses on the 
central charge of the Virasoro algebra in (\ref{algescal}), in such
a way that they vanish in the classical limit $c\rightarrow\infty$ 
(see \cite{Witten}), thus revealing themselves as a quantum 
phenomenon.
Let us finally state that, though the 
presence of (non-gravitational) interactions
is crucial in ascribing a point-like nature to the particles in QFT, once we 
accept such a nature we can get rid of those interactions
and deal essentially with {\it free} theories as a starting point for the
proposed mechanism.
Therefore, even if our conjecture of relating a mass origin
for particles to the interaction with the some effective 
degrees of freedom does not fully work, the proposed radiative corrections
could imply important and observable consequences on the energy spectrum of 
{\it free} fields, in particular entailing modifications in the field 
propagators.

\nonumsection{Acknowledgments}
\noindent
We want to thank C. Barcel\'o for crucial discussions and his
continuous support. We also want to thank A.P. Balachandran for stimulating 
comments.

\nonumsection{References}

\end{document}